%% file: error.tex
\documentclass[a4paper,11pt]{scrartcl}
\usepackage[centertags]{amsmath}
\usepackage{amsfonts}
\usepackage{amssymb}
\usepackage{amsthm}
\usepackage{pifont}
\usepackage[ansinew]{inputenc}
\usepackage{graphicx}
\usepackage{natbib}

\begin{document}

\title{On the error of incidence estimation from prevalence data}

\author{Ralph Brinks\footnote{ralph.brinks@ddz.uni-duesseldorf.de}\\
Institute for Biometry and Epidemiology\\German Diabetes Center\\
Düsseldorf, Germany}

\date{}

\maketitle

\begin{abstract}
This paper describes types of errors arising in a recently proposed
method of incidence estimation from prevalence data. The errors are illustrated by
a simulation study about a hypothetical irreversible disease. In addition, a way of
obtaining error bounds in practical applications of the method is proposed.
\end{abstract}


\input{error2_10}

\end{document}

%% file: error2_10.tex
\section{Introduction}
Recently, we have shown how to estimate the incidence of an irreversible disease
by the age-specific prevalence in case the mortality of the diseased and the healthy population
are known, \citep{Bri13}. The age-specific prevalence can, for instance, be obtained 
from cross-sectional studies.
In \citep{Bri13} one cross-section was used to estimate the incidence of renal failure.
The underlying approach had been an ordinary differential equation (ODE) which is valid 
if the only relevant time-scale is the age of the persons in the considered population. 
Error considerations have been treated by a bootstrap approach. 

Later we have proven that the underlying ODE 
is a special case of a partial differential equation (PDE) that involves additional time scales,
\citep{Bri13a} and \citep{Bri14}. If we want to use the PDE for estimation of the incidence, at 
least two cross-sections are necessary, \citep{Bri13b}. 

This work deals with incidence estimation from two cross-sections using the PDE approach. It is
shown that the incidence estimation is affected by three types of error: (i) 
a systematic error which is given by the study design (or the available data), 
(ii) the sampling of the age course, and 
(iii) by the error attributable to sampling the population. After a short summary of our previous 
works alongside with an
introduction of the notation in this article, an example is presented and
the types of error are introduced and examined.

\section{Illness-death model} 
\begin{figure*}[ht!]
\centerline{\includegraphics[keepaspectratio,
width=10cm]{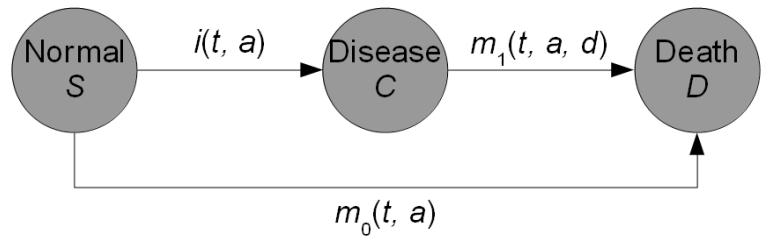}} \caption{Illness-death model 
of an irreversible disease. The
transition rates between the three states depend on the calendar time 
$t,$ on the age $a,$ and
in case of the disease-specific mortality $m_1$ also on the
disease's duration $d$.} \label{fig:3states}
\end{figure*}

In dealing with the incidence, prevalence and mortality with respect to a disease, 
it is useful to look at the illness-death model shown in Figure \ref{fig:3states}, \citep{Kal02}. 
The transition rates  are the incidence rate $i,$ and $m_0$ and $m_1$ are the mortality rates
of the non-diseased 
and diseased persons, respectively. In general, these rates depend on calendar time 
$t$, age $a$ and in case of $m_1$ also on the duration $d$ of the disease.

In \citep{Bri13a} it has been shown that the age-specific prevalence\footnote{The 
number of diseased persons aged $a$ at time $t$ over the total number of living persons age $a$ at $t.$}
$p(t, a)$ of the disease is related to the rates $i$, $m_0$ and $m_1$ by following PDE:

\begin{equation}\label{e:brinks}
\left ( \frac{\partial}{\partial a} + \frac{\partial}{\partial a} \right ) p = 
(1-p) \cdot \Bigl ( i - p \cdot \left (m^\star_1 -
m_0\right ) \Bigr ).
\end{equation}

In Equation \eqref{e:brinks} $m_1^\star(t, a)$ is the \emph{overall
mortality rate}. The overall mortality $m_1^\star$ is the mortality that would be surveyed in a
representative sample of the diseased population. As shown in \citep{Bri13a}, it can be expressed by
\begin{equation}\label{e:mstar2}
 m_1^\star(t, a) = \frac{\int\limits_0^a m_1(t, a, \delta) \, i(t-\delta, a-\delta) \, 
               \mathcal{M}_{t,a}(a-\delta) \,
          e^{- M_1(t, a, \delta)} \, \mathrm{d}\delta}
                       {\int\limits_0^a i(t-\delta, a-\delta) \, \mathcal{M}_{t,a}(a-\delta) \,
          e^{- M_1(t, a, \delta)} \, \mathrm{d}\delta},
\end{equation}
where
\begin{equation*}
\mathcal{M}_{t, a}(y) := \exp \left ( -\int_0^y m_0(t-a+\tau,
\tau) + i(t-a+\tau, \tau) \mathrm{d}\tau \right )
\end{equation*}
and
$$M_1(t, a, d) := \int_{0}^{d} m_1(t - d + \tau, \,a - d + \tau , \,\tau ) 
\, \mathrm{d}\tau.$$

\section{Simulation study: incidence by two two cross-sections} 
Consider an hypothetical irreversible disease, whose incidence we want to estimate
from two cross-sectional studies at two different points $t_k, ~k=1, 2,$ in time. Let be 
$t_1 < t_2.$ The
outcomes of the cross-sectional studies are the age-specific prevalences $p(t_k, \cdot), ~k=1, 2.$ 
We want to use Equation \eqref{e:brinks}, which requires the approximation of the partial
derivative $(\tfrac{\partial}{\partial a} + \tfrac{\partial}{\partial a}) p$ from the $p(t_k, \cdot), 
~k=1, 2.$ Thus, it is reasonable to estimate the incidence in the middle $t_s = \tfrac{1}{2}(t_1+t_2)$
of the interval $[t_1, ~t_2].$ For our example we assume to know the age-specific mortality rates $m_0$ and 
$m_1^\star$ at $t_s.$ We set up a simulation study to analyse the performance of the incidence estimation.

\bigskip

For our simulation, we consider a population moving in the illness-death model very much alike
as described in \citep[Simulation 2]{Bri14a}. We mimic two cross-sections at $t_1 = 100$ 
and $t_2 = 110,$ and estimate the age-specific incidence $i$ at $t_s = 105.$ 
Since we know the true incidence
underlying the simulation, we can compare the estimate with the true incidence.

As in \citep[Simulation 2]{Bri14a}, the incidence of a hypothetical disease is assumed 
to be $i(t, a) = (a-30)_+/3000.$ Here, the notation $x_+$ means
$x_+ = \max(0, x).$ The
mortality of the non-diseased is $m_0(t,a) = \exp(-10.7 + 0.1 a)
\cdot 0.998^t$ and the mortality of the diseased population is $m_1(t, a, d)
= m_0(t, a) \cdot 0.04 (d - 5)^2 + 1.$ With these information we can compute $m_1^\star(t_s, a)$ 
by Equation \eqref{e:mstar2}. This is done by Romberg integration with a prescribed accuracy, 
\citep{Dah74}.

\subsection{Systematic error due to study design} 
Based on the information about $i, m_0$ and $m_1^\star$ we can \emph{calculate} the prevalence 
$p(t_k, \cdot), ~k=1, 2,$ by numerically solving Equation \eqref{e:brinks}. Alternatively, we
can apply Keiding's formula \citeyearpar[Section 7.2]{Kei91}, which in our notation reads as
\begin{equation}\label{e:Keiding}
p(t, a) = \frac{\int\limits_0^a i(t-\delta, a-\delta) \, \mathcal{M}_{t,a}(a-\delta) \,
          e^{- M_1(t, a, \delta)} \, \mathrm{d}\delta}
          {\mathcal{M}_{t,a}(a) + \int\limits_0^a i(t-\delta, a-\delta) \, \mathcal{M}_{t,a}(a-\delta) \,
          e^{- M_1(t, a, \delta)} \, \mathrm{d}\delta}.
\end{equation}

\begin{figure*}[ht!]
\centerline{\includegraphics[keepaspectratio,
width=14cm]{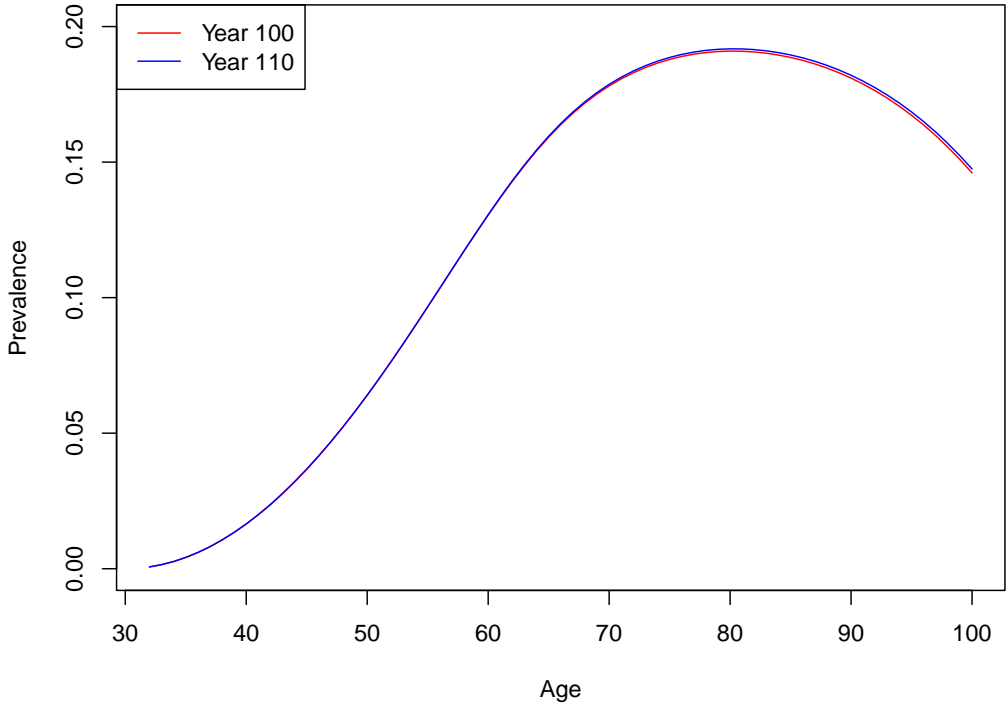}} \caption{Age-specific prevalence of an hypothetical irreversible
disease at times $t_1 = 100$ (red) and $t_2 = 110$ (blue).} \label{fig:prev}
\end{figure*}

The result of Romberg-integrating Equation \eqref{e:Keiding} is shown in Figure \ref{fig:prev}.
The age courses of the prevalence in $t_1$ and $t_2$ differ only slightly. 

Both curves in Figure \ref{fig:prev}
are ideal in the sense that no error due to sampling occurs, they are exact (within 
the prescribed error bounds resulting from the Romberg integration). Before we study the effects
of sampling errors, we try to \emph{reconstruct} the incidence from these ideal curves. The
term \emph{reconstruction} is deliberately chosen to contrast it against the term \emph{estimate}, 
which is used later and involves a sampling component.

To reconstruct $i(t_s, \cdot)$ from the $p(t_k, \cdot), ~k=1,2,$ we solve Equation \eqref{e:brinks}
for $i:$

\begin{equation}\label{e:solved4i}
    i(t_s, a) = \frac{(\partial_t + \partial_a) p (t_s, a)}{1-p(t_s, a)}
                                          + p(t_s, a) \; \Bigl( m^\star_1 (t_s, a) - m_0(t_s, a) \Bigr ).
\end{equation}

For ease of notation, we have written $\partial_x = \tfrac{\partial}{\partial x}$ for $x \in \{t, a\}.$
Note that we assume $m^\star_1 (t_s, a)$ and $m_0(t_s, a)$ to be known for all $a.$ From the remaining
quantities in Equation \eqref{e:solved4i} the prevalence $p(t_s, \cdot)$ and the partial derivative 
$(\partial_t + \partial_a) p (t_s, \cdot)$ are unknown. We use following approximations:

\begin{equation}\label{e:intermediateP}
    p(t_s, a) \doteq \tfrac{1}{2} \; \left [ p(t_1, a) +  p(t_2, a) \right ].
 \end{equation}
and
\begin{equation}\label{e:approxDP}
   (\partial_t + \partial_a) p (t_s, a) \doteq \tfrac{1}{\Delta} \; 
           \left [ p(t_1, a + \tfrac{\Delta}{2}) - 
                   p(t_2, a - \tfrac{\Delta}{2}) \right ],
\end{equation}
with $\Delta = t_2 - t_1.$

\begin{figure*}[ht!]
\centerline{\includegraphics[keepaspectratio,
width=14cm]{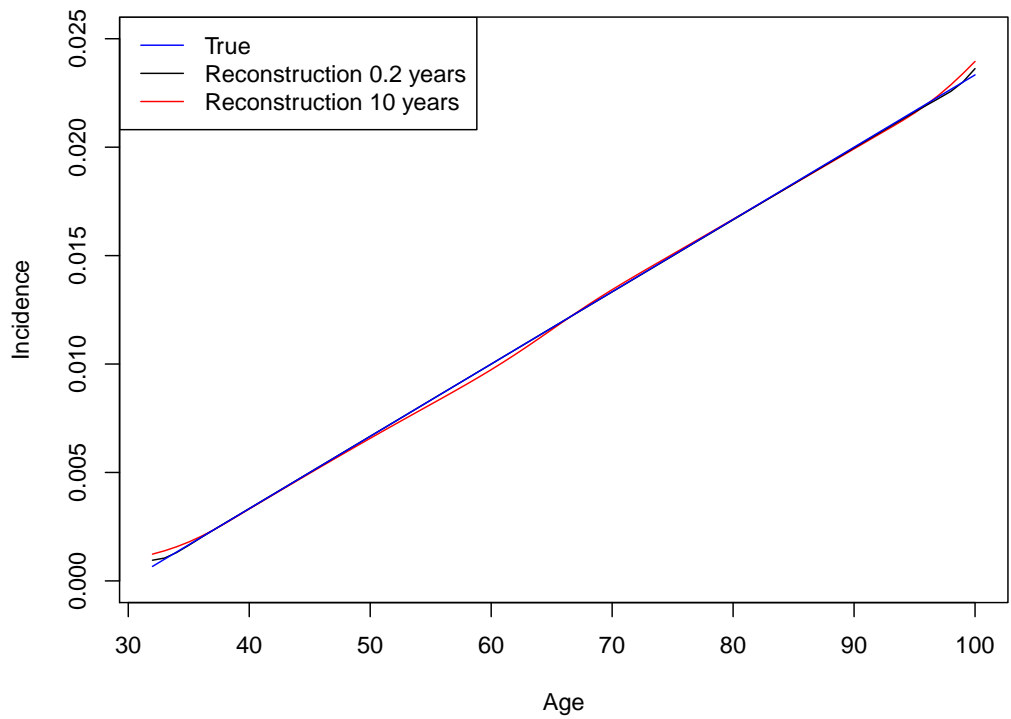}} \caption{True (blue) and reconstructed 
age-specific incidences for $\Delta = 10$ (red) and $\Delta = 0.2$ (black). The black line
coincides very well with the blue line.} \label{fig:desErr}
\end{figure*}

If we use Equation \eqref{e:solved4i} with the approximations \eqref{e:intermediateP} and \eqref{e:approxDP}
we obtain the reconstructed incidence as shown in Figure \ref{fig:desErr}. The blue line represents the true
incidence, the red line the reconstructed incidence. We can see slight differences between these lines.
The relative differences (in \%) between the reconstructed and the true incidence is shown in Table \ref{t:dev1}.
We can see that the greatest relative deviation occurs at the lowest (7.68\% at age 35) and the highest 
age class (2.66\% at age 100). Since these deviations are not attributable to sampling error but just to the
approximations \eqref{e:intermediateP} and \eqref{e:approxDP} and thus by the choice of $t_1$ and $t_2,$ 
we call these \emph{errors by study design.} These errors are intrinsic to the choice of $t_k, ~k=1,2,$ which
in an epidemiological application are given by the available data.

We can see that indeed the study design is responsible for the relatively high deviations. If we choose $t_1 = 104.9$
and $t_2 = 105.1,$ we can see that the deviation between the true and the reconstructed
incidence decreases. The reconstructed incidence for $\Delta = 0.2$ is depicted as a black line 
in Figure \ref{fig:desErr}. It is closer to the true incidence than the red line. The
relative error for this case is shown in the third column of Table \ref{t:dev1}.

\begin{table}[ht]
\centering
\begin{tabular}{ccc}
  \hline
Age & Rel. error (\%) & Rel. error (\%) \\ 
(in years) & $\Delta = 10$ & $\Delta = 0.2$ \\ 
  \hline
35.0& 7.68 &  -1.48 \\ 
40.0& -0.56 &  0.01 \\ 
45.0& -0.75 & -0.00 \\ 
50.0& -1.38 &  0.01 \\ 
55.0& -2.36 &  0.01 \\ 
60.0& -2.61 & -0.02 \\ 
65.0& -0.70 &  0.00 \\ 
70.0&  0.72 &  0.01 \\ 
75.0&  0.43 & -0.00 \\ 
80.0&  0.06 &  0.00 \\ 
85.0& -0.20 & -0.00 \\ 
90.0& -0.39 &  0.01 \\ 
95.0& -0.42 & -0.00 \\ 
100.0& 2.66 &  1.54 \\    \hline
\end{tabular}
\caption{Relative errors of the reconstructed incidences at specific ages. The second and third column
shows the relative errors for the inital setting with $\Delta = 10$ and the 
decreased relative errors for $\Delta = 0.2.$}\label{t:dev1}
\end{table}

It is important to remember that the deviations described so far are just caused by the choice of
the study design, not by any sampling uncertainty. Since we use the approximations in 
Equations \eqref{e:intermediateP} and \eqref{e:approxDP}, which are exact only in special
cases, we will almost always have an error in the reconstructed incidence.

\clearpage

\subsection{Sampling error of the prevalence} 
In the previous section we have used exact values for $p(t_k, \cdot), ~k=1,2.$ Surveying prevalence
in real cross-sectional studies usually suffer from several sources of errors. Examples are
measurement errors (i.e., errors in determining the state a subject belongs to in the illness-death 
model), non-representativeness of the study participants (selection bias), discretisation error 
and sampling error. We confine ourselves to the last types of errors. 

\subsubsection{Error types}
By \emph{discretisation error} we mean everything that is related to making the continuous functions 
$a \mapsto p(t_k, a), ~k=1,2,$ \emph{discrete}. Typically the prevalence 
$p(t_k, \cdot)$ is estimated using finitely many age groups. 
Assumed we want to estimate $p(t_k, \cdot)$ at ages 
$a_\ell, ~\ell = 1, \dots, L,$ then all persons alive at $t_k$ whose
age is in the age group $(a_\ell - \varepsilon, a_\ell + \varepsilon], ~\varepsilon > 0,$ are examined 
if the have the disease or not. Following approximation is used: 

\begin{equation*}
p(t_k, a_\ell) \approx 
\frac{\# \text{persons alive at } t_k \text{ having the disease and age in } (a_\ell - \varepsilon, a_\ell + \varepsilon]}
     {\# \text{persons alive at } t_k \text{ with age in } (a_\ell - \varepsilon, a_\ell + \varepsilon]}.
\end{equation*}

Situations are easily imaginable, where this estimation is biased, for instance, if $a_\ell$ is a
local extremum of $p(t_k, \cdot).$

\bigskip

Finally, by \emph{sampling error} we mean any effect that is related to having only a sample of the whole 
population in the study. 

\subsubsection{Sampling error}
To examine the effect of the sampling error, we simulate a population in the illness-death model.
Each individual is disease-free at birth and is followed from birth to death (without loss). 
In each of 70 consecutive years $t = 1, \dots, 70,$ we
consider 300000 persons born with date of birth uniformly
distributed across the year. In total 21 million ($=70 \times 300000$) persons are simulated.
Incidence $i$ and $m_0$ are treated as competing risk, and details of
the implementation (with source code) are described in \citep{Bri14a}. 

The cross-sections in the years $t_1 = 100$ and $t_2 = 110$ comprise more than 
11 million persons alive aged between 40 and 95. The age distribution of the living and 
the prevalent persons
are shown in Table \ref{t:cross1}.

\begin{table}[ht]
\centering
\begin{tabular}{c|rr|rr}
  \hline
Age     &  \multicolumn{4}{c}{Cross-section at}  \\
group   & \multicolumn{2}{|c}{$t_1 = 100$} & \multicolumn{2}{|c}{$t_2 = 110$} \\
(years) & alive (N) & prevalent (n) & alive (N) & prevalent (n) \\ 
  \hline
(40,45] & 1479755 &  38605 & 1480777 &  38391 \\ 
(45,50] & 1467357 &  72794 & 1468037 &  73006 \\ 
(50,55] & 1445529 & 115689 & 1445263 & 115545 \\ 
(55,60] & 1405534 & 159514 & 1407035 & 159795 \\ 
(60,65] & 1333192 & 194120 & 1336987 & 194515 \\ 
(65,70] & 1215229 & 205780 & 1218949 & 206676 \\ 
(70,75] & 1041215 & 190407 & 1046718 & 192903 \\ 
(75,80] &  819662 & 155683 &  828053 & 157549 \\ 
(80,85] &  568871 & 108209 &  577779 & 110667 \\ 
(85,90] &  326252 &  60093 &  332622 &  61667 \\ 
(90,95] &  137577 &  24146 &  143004 &  25145 \\ 
   \hline
(40, 95]&11240173 &1325040 &11285224 & 1335859 \\
   \hline
\end{tabular}
\caption{Age-distributions of living and prevalent persons in the two cross-sections.}\label{t:cross1}
\end{table}

\begin{figure*}[ht]
\centerline{\includegraphics[keepaspectratio,
width=14cm]{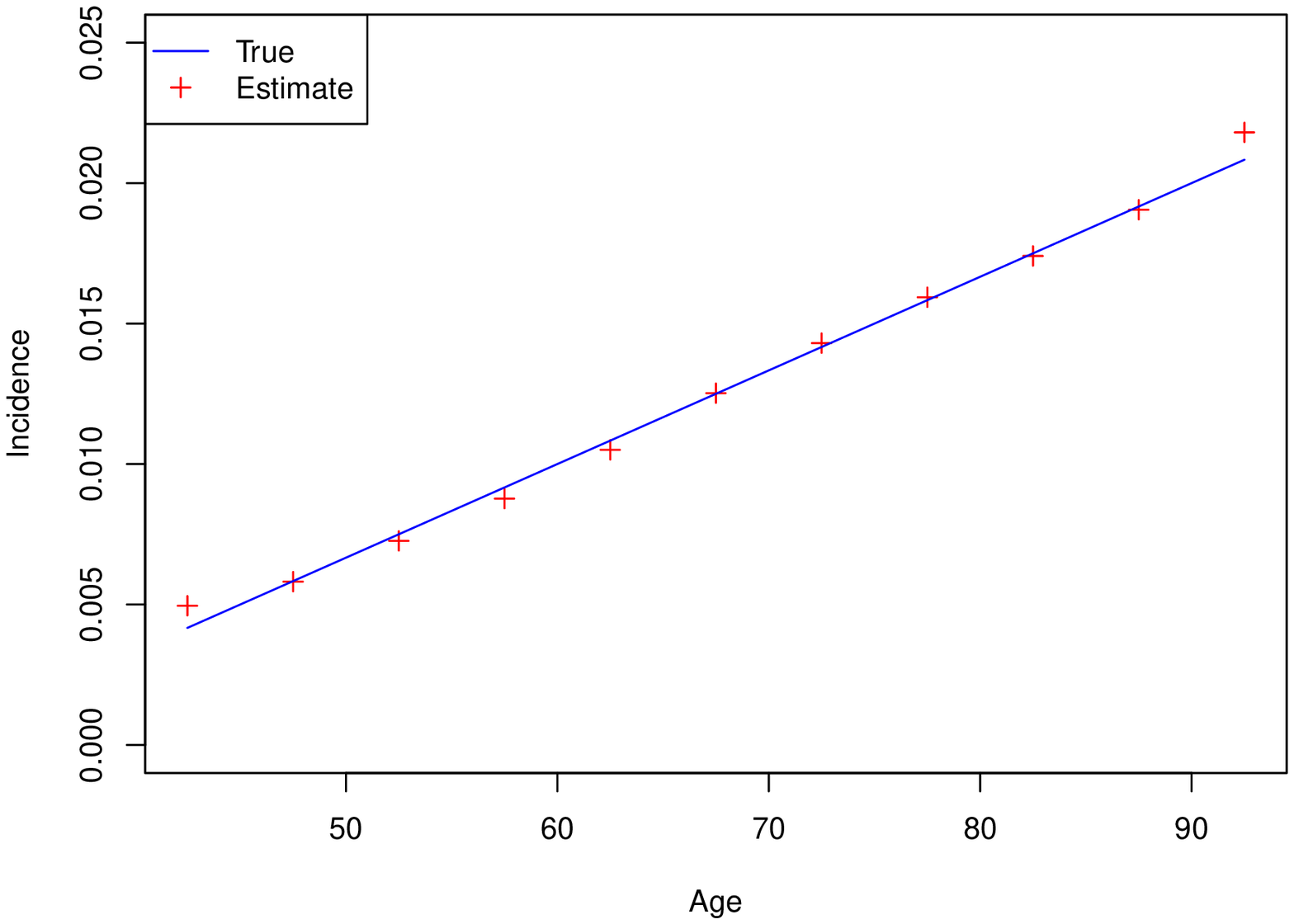}} \caption{True (blue) and estimated (red)
age-specific incidences based on the age-specific prevalences from Table \ref{t:cross1}.} 
\label{fig:discreteErr}
\end{figure*}

From the resulting age-specific prevalence via Equation \eqref{e:solved4i} with the approximations 
\eqref{e:intermediateP} and \eqref{e:approxDP} we obtain the estimated age-specific incidence 
$i(t_s, a_\ell), ~\ell = 1, \dots, L,$ as shown in Figure \ref{fig:discreteErr}. The error due
to the study design is still visible (cf. Figure \ref{fig:desErr}). Table \ref{t:dev2} shows 
the relative errors of the estimated age-specific incidence.

\begin{table}[ht]
\centering
\begin{tabular}{cc}
  \hline
 $a_\ell$ & Rel. error (\%) \\ 
  \hline
42.5 & 18.89 \\ 
47.5 & -0.35 \\ 
52.5 & -3.14 \\ 
57.5 & -4.33 \\ 
62.5 & -3.03 \\ 
67.5 &  0.18 \\ 
72.5 &  0.97 \\ 
77.5 &  0.65 \\ 
82.5 & -0.53 \\ 
87.5 & -0.59 \\ 
92.5 &  4.67 \\ 
   \hline
\end{tabular}
\caption{Relative errors of the estimated age-specific incidence based on 
the data in Table \ref{t:cross1}. The numerical values of the estimated
incidence are shown in the fourth column of Table \ref{t:incEstimates}.}\label{t:dev2}
\end{table}

\clearpage

In the next step, we want to study the impact of including a lower number of persons in the cross-sectional
studies at $t_1$ and $t_2.$ For this, we repetitively ($n_\text{rep} = 1000$) draw samples of different
sizes ($N_\kappa$) from the population of the 21 million, estimate the incidence for $a_\ell$ in 
the described way and examine the distribution of the $n_\text{rep}$ estimates of the incidence.

\begin{figure*}[ht!]
\centerline{\includegraphics[keepaspectratio,
width=14cm]{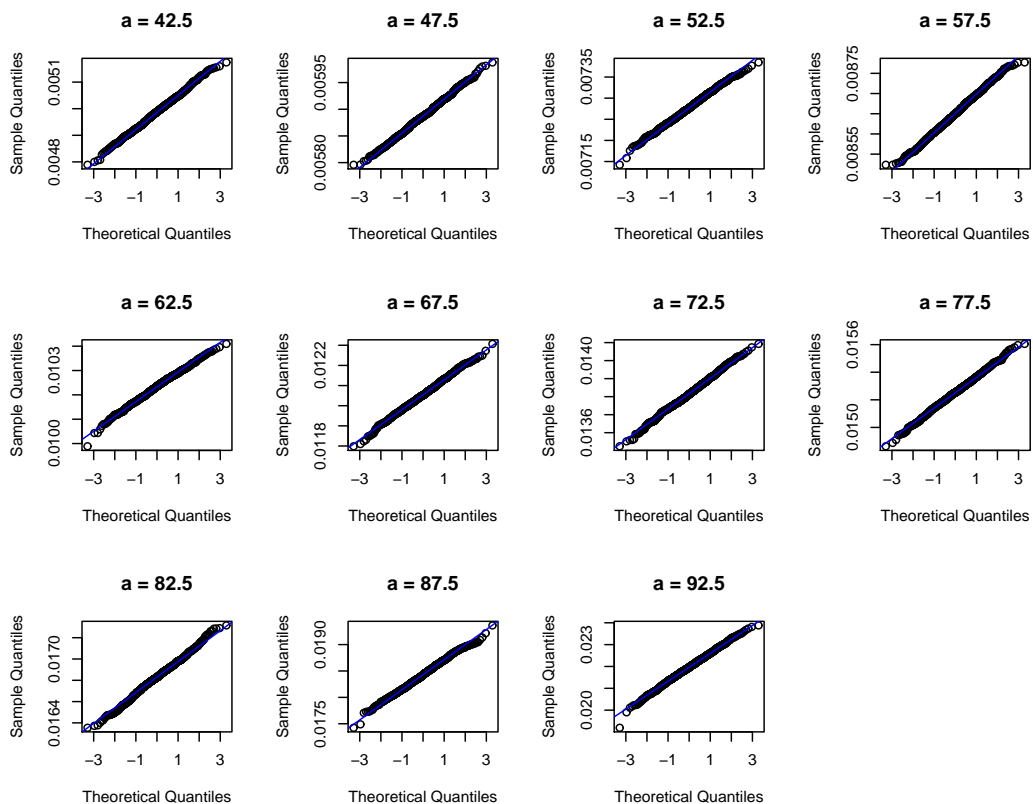}} \caption{Q-Q-plots of the age-specific incidence estimates
based on $n_\text{rep} = 1000$ subpopulations (of size $N_1 = 2.6$ million) drawn from the 
original population of size 21 million. We may conclude that in all age groups the 
estimates are normally distributed.} 
\label{fig:qq}
\end{figure*}

Figure \ref{fig:qq} shows the quantile-quantile plots (Q-Q-plots) of the $n_\text{rep}$ repeated estimates for a
subpopulation of size $N_1 = 2.6$ million for the different age groups $a_\ell$ compared to the normal
distribution. In all age groups $a_\ell$ the Q-Q-plots indicate that the estimates are normally distributed.
For $N_\kappa, ~\kappa = 2, 3,$ these Q-Q-plots (not shown here) allow the same conclusion.

Based on the observation that the incidence estimates follow a normal distribution 
(see Figure \ref{fig:qq}), the distribution may be characterised by mean and standard 
deviation (SD). The fifth to the tenth column of Table \ref{t:incEstimates} show
the corresponding values for $N_1 = 2.6$ million, $N_2 = 650000$ and $N_3 = 130000.$
We can see that the mean of the distribution remains stable whereas the SD approximately
doubles in each step from left to right. The doubling of the SD is not surprising as 
$\tfrac{N_{\kappa}}{N_{\kappa+1}} \approx 4, ~\kappa = 1,2.$

\begin{table}[ht]
\centering
\begin{tabular}{c|c|c|c|cc|cc|cc}
  \hline
Age      & True      & Without  &  Population    & \multicolumn{2}{|c}{$N_1 = 2.6$ mio} & 
                   \multicolumn{2}{|c}{$N_2 = 650000$} & \multicolumn{2}{|c}{$N_3 = 130000$} \\ 
$a_\ell$ & incidence & sampling &  of 21 mio.& Mean & SD & Mean & SD & Mean & SD \\ 
  \hline
42.5 &  41.7 &  41.4 &  49.5 &  49.9 & 0.9 &  49.9 &  1.8 &  49.9 &  3.5\\ 
47.5 &  58.3 &  57.7 &  58.1 &  58.8 & 0.4 &  58.8 &  0.9 &  58.9 &  1.9\\ 
52.5 &  75.0 &  73.6 &  72.6 &  72.8 & 0.5 &  72.8 &  1.0 &  72.6 &  2.0\\ 
57.5 &  91.7 &  89.2 &  87.7 &  86.5 & 0.6 &  86.5 &  1.3 &  86.3 &  2.6\\ 
62.5 & 108.3 & 106.3 & 105.1 & 102.4 & 0.8 & 102.3 &  1.6 & 102.5 &  3.3\\ 
67.5 & 125.0 & 125.4 & 125.2 & 120.5 & 1.0 & 120.5 &  2.1 & 120.6 &  4.2\\ 
72.5 & 141.7 & 142.7 & 143.0 & 138.1 & 1.3 & 138.1 &  2.5 & 137.9 &  5.1\\ 
77.5 & 158.3 & 158.9 & 159.4 & 152.5 & 1.7 & 152.4 &  3.3 & 152.2 &  6.6\\ 
82.5 & 175.0 & 175.1 & 174.1 & 168.3 & 2.1 & 168.2 &  4.4 & 168.0 &  8.9\\ 
87.5 & 191.7 & 191.4 & 190.5 & 184.4 & 3.8 & 184.6 &  7.7 & 184.4 & 15.4\\ 
92.5 & 208.3 & 207.7 & 218.1 & 219.8 & 9.3 & 220.0 & 18.7 & 221.7 & 37.9\\ 
   \hline
\end{tabular}
\caption{True and estimated incidences in the different age groups based on different
prevalence data.}\label{t:incEstimates}
\end{table}

\begin{figure*}[ht]
\centerline{\includegraphics[keepaspectratio,
width=14cm]{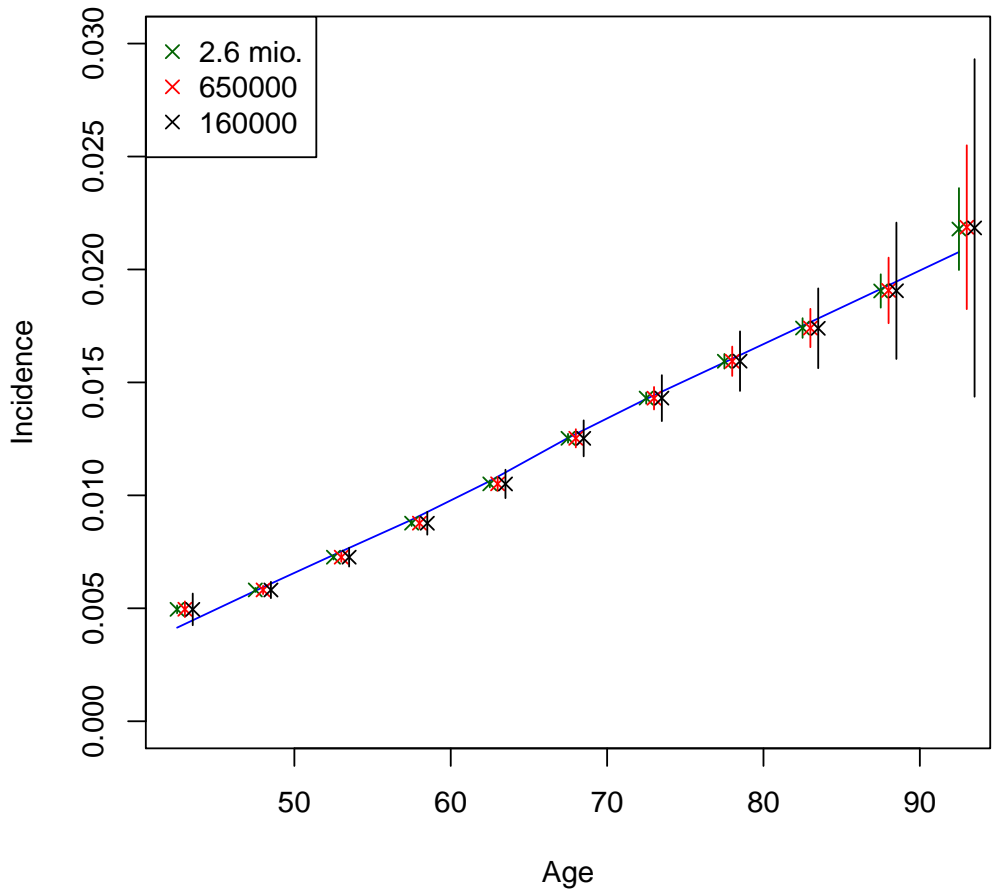} } 
\caption{Mean vales and the central 95\% coverage intervals of the estimated 
incidences based on the different population sizes $N_\kappa, ~\kappa = 1, 2, 3.$} 
\label{fig:sampleError}
\end{figure*}

In Figure \ref{fig:sampleError} the mean and the 95\% coverage intervals of the estimated 
incidences based on the different $N_\kappa, ~\kappa = 1, 2, 3,$ are shown. The 
design error is still visible in the point estimates (cf. Figure \ref{fig:desErr}). 

\clearpage

\section{Discussion} 
In this work we examine different sources of errors in a recently proposed
algorithm of estimating incidences from two cross-sections. The first source of 
error is due to the study design. Two cross-sections at different points 
in time $t_k, ~k=1, 2,$ can only approximate the partial
derivative of the prevalence. This error is intrinsic of the data sources 
available. In our example, a smaller difference between the $t_k$ is favourable
over a larger difference (Table \ref{t:dev1}). In practice, this may not always be the
case.
The second source of error, the discretisation error, is a result from estimating the 
prevalence at a specific age by the prevalence in an age group. Situations are
possible, where the prevalence at a specific age is not accurately estimated by
the prevalence in an age group.
Finally, sampling error due to the limited persons in the cross-sectional studies
is examined. It can be seen that imprecise estimates of the prevalence due to
few persons in the age groups leads to inaccurate estimates of the incidence.

This work just examines the impact of errors in the prevalence due to study design, 
discretisation and sampling. In epidemiological applications the estimates of the mortality
rates are also subject to errors. These are not considered here. However, this work sketches an
easy way to obtain error bounds of incidence estimates in this context as well: Based on the 
uncertainties in the input values, prevalence and mortality rates, one may draw
random samples from the distributions of input values, apply the framework shown in this article 
and then examine the distributions of the estimated incidences.